# Identifying Barriers Hindering the Acceptance of Generative AI as a Work Associate, measured with the new AGAWA scale

Łukasz Sikorski[a] luk_sikorski@umk.pl - corresponding author, ORCID: 0009-0009-6564-6594

Albert Łukasik[b] ORCID: 0000-0002-7283-7999

Jacek Matulewski[a] jacekmatulewski@umk.pl, ORCID: 0000-0002-1283-6767

Arkadiusz Gut[c], ORCID: 0000-0002-3983-9474

[a] Department of Informatics, Faculty of Physics, Astronomy and Informatics, Nicolaus Copernicus University in Toruń, Grudziądzka 5, 87-100 Toruń, Poland

[b] Department of Cognitive Science, Doctoral School of Social Sciences, Nicolaus Copernicus University in Toruń, Fosa Staromiejska 1a, 87-100 Toruń, Poland

[c] Department of Cognitive Science, Faculty of Philosophy and Social Sciences, Nicolaus Copernicus University in Toruń, Fosa Staromiejska 1a, 87-100 Toruń, Poland

**Acknowledgments**
We would like to thank Oleg Gorbaniuk for sharing his knowledge of statistical analysis.

**Declaration of competing interests**
The authors declared no potential conflicts of interest with respect to the research, authorship, and/or publication of this article.

## Abstract

The attitudes of today's students toward generative AI (GenAI) will significantly influence its adoption in the workplace in the years to come, carrying both economic and social implications. It is therefore crucial to study this phenomenon now and identify obstacles for the successful implementation of GenAI in the workplace, using tools that keep pace with its rapid evolution. For this purpose, we propose the AGAWA scale, which measures attitudes toward an artificial agent utilising GenAI and perceived as a coworker. It is partially based on the TAM and UTAUT models of technology acceptance, taking into account issues that are particularly important in the context of the AI revolution, namely acceptance of its presence and social influence (e.g., as an assistant or even a supervisor), and above all, resolution of moral dilemmas. The advantage of the AGAWA scale is that it takes little time to complete and analyze, as it contains only four items.

In the context of such cooperation, we investigated the importance of three factors: concerns about interaction with GenAI, its human-like characteristics, and a sense of human uniqueness, or even superiority over GenAI. An observed manifestation of the attitude towards this technology is the actual need to get help from it. The results showed that positive attitudes toward GenAI as a coworker were strongly associated with all three factors (negative correlation), and those factors were also related to each other (positive correlation). This confirmed the relationship between affective and moral dimensions of trust towards AI and attitudes towards generative AI at the workplace.

# Introduction

## Barriers to implementing GenAI as a work associate

Generative artificial intelligence (GenAI) has proven to be a key element of the phenomenon known as Industry 4.0. It is fundamentally changing how we work, how we manage work, and the labor market itself, replacing some professions and creating others (WEF, 2025; Brynjolfsson et al., 2025; Salari et al., 2025; Law & Varanasi, 2025), just as previously did the mechanization of production, the introduction of mass production, and the spread of computers. These changes have profound social implications, which means that they are accompanied by concerns about maintaining employment, naturally leading to resistance to adapting to the new technology, as is currently the case with GenAI in the workplace

(Schwab, 2016). Similar to the three previous industrial revolutions, while increasing the capabilities of humans as a species, it also pushes the boundaries of their uniqueness and autonomy. The attitude of employees towards GenAI will have a decisive impact on the success of this process.

Attitudes toward artificial intelligence (AI) vary among the general population and are influenced by several factors. Research indicates that they are shaped by the level of knowledge about the technology (Kaya et al., 2024; Milicevic et al., 2024) and prior experience with its use (Cardon et al., 2023; Lovato et al., 2024). Age (Syzygy, 2017; Gillespie et al., 2021; Lovato et al., 2024), gender (Bergdahl et al., 2023; Liang and Lee, 2017; Sindermann et al., 2022), education level (Gillespie et al., 2021), and income (Bergdahl et al., 2023) are also important elements. Additional factors include occupation (Boucher et al., 2024; Sikorski et al., 2025; GDC, 2024) and the purpose for which AI is used (Gillespie et al., 2021; Schepman and Rodway, 2020).

Although general attitudes toward AI tend currently to be positive (Schepman and Rodway, 2020; Gillespie et al., 2021), certain factors still distort its reception. A key and frequently mentioned concern is competition in the labor market and the fear of job loss resulting from widespread and rapid AI implementation (Schepman and Rodway, 2020; Hopelab et al., 2024; Almaraz-López et al., 2023). It is not surprising, as AI tools significantly enhance work efficiency (Karamthulla et al., 2024), which has already led to a reduction in the recruitment of university graduates (SignalFire, 2025). According to a World Economic Forum report (2025), 40% of employers plan to reduce headcount due to automation enabled by AI. Even now, this technology is already widely adopted. In 2024, 75% of global knowledge workers, including 85% of those identified as Gen Z, reported using AI in their jobs (Microsoft and LinkedIn, 2024). A McKinsey report (2024) also showed that 66% of surveyed companies used AI tools. Analyzing the labor market of the entire world, it is estimated that 24% of workers are employed in occupations with some "GenAI exposure", and 3.3% in occupations that can be automated with generative AI (GenAI) to the highest degree (Gmyrek et al., 2025). In 2024, approximately 3.5 billion people were employed globally (International Labour Organization, 2024), so GenAI will affect around 840 million employees. Interestingly, concerns about labor market competition seem not to correlate directly with overall levels of fear toward AI-based technologies (Kaya et al., 2024).

AI's influence on the modern world also generates anxiety among students and affects their choice of field of study. For instance, medical students are less likely to choose specific specializations they fear may be overtaken by AI, such as radiology (Park et al., 2021; Dahmash et al., 2020). Similarly, students in business and economics express concerns that AI may dominate specific career paths related to their field shortly (Almaraz-López et al., 2023). Additionally, computer science students fear competition from generative AI (GenAI) more than professional software developers (Stack Overflow, 2024).

Apart from concerns about competition with GenAI in the job market, ambivalent attitudes toward this technology are also rooted in moral concerns. These include uncertainty around ownership of training content and copyright issues related to AI-generated outputs (Vimpari et al., 2023). Ethical concerns also arise from the risk of bias and discrimination stemming from the quality of training data (Karamthulla et al., 2024). Additional worries include the potential devaluation of creative labor (Vimpari et al., 2023) and increased use of personal data by AI-driven applications (Schepman and Rodway, 2020).

## Factors investigated in this study

One of the considered obstacles is the fear of artificial agents. For robotic agents, a similar construct is measured by the *Negative Attitudes toward Robots Scale* (NARS) (Nomura et al., 2006), which includes two sub-scales: the NATIR, which measures the fear of interacting with robots, and the NARHT, which measures the fear of robots with human characteristics. Originally developed in Japanese, the scale has been translated and validated for Portuguese (Piçarra et al., 2015) and Polish populations (Pochwatko et al., 2015). The latter adaptation is crucial in the context of this research, as it was conducted on the Polish population. Importantly, it is open to adaptation for different artificial agents and indeed has often been modified to assess aversion and discomfort in interactions with avatars (Diana et al., 2022; Kodani et al., 2024), chatbots (Yang et al., 2022), and AI (Park and Kim, 2024).

Alongside the NARS questionnaire, the *Belief in Human Nature and Uniqueness Scale* (BHNU) is frequently used in studies (Giger et al., 2025; Łupkowski and Gierszewska, 2019) on attitudes toward artificial intelligence, which measures the third factor taken into account, namely the feeling of superiority over artificial agents. BHNU serves as a socio-cognitive predictor of beliefs concerning the distinctiveness of humans in relation to machines (Giger et al., 2017) and avatars (Ratajczyk et al., 2024; Stein et al., 2019). These beliefs regarding the uniqueness of human consciousness, morality, and emotionality seem to form an essential framework for AI attitudes.

## Construct of Attitudes toward GenAI

A positive attitude toward GenAI as a potential workplace associate includes:
1) the belief that cooperation, understood as engaging in interaction with a GenAI agent, positively affects professional performance (i.e., the usefulness of this technology and the benefits of using it), and
2) general acceptance of collaboration with GenAI, including the belief that it does not raise moral concerns.

In contrast, a negative attitude refers to the refusal to acknowledge GenAI's effectiveness or concerns about its use, including both
1) moral objections and
2) fears of its negative social impact, implicitly regarding, among others, the labor market.

These beliefs and fears relate to general evaluations and participants' individual assessments of their professional circumstances.

The above attitudes toward GenAI can be treated as a concretization of the more generally understood attitude toward technology described in the Technology Acceptance Model theory, however, tailored to the specific characteristics of GenAI and consequences of its implementation, which are already widely visible. Unlike earlier technologies, generative AI not only automates tasks but also produces creative and cognitive outputs, raising novel challenges for how employees perceive their own roles, competencies, and future prospects (Granulo et al., 2019).

# Measuring Attitude Towards AI

Examining attitudes toward AI has become particularly important (Bergdahl et al., 2023; Latikka et al., 2023), covering aspects such as trust, readiness to use, fear, and perceived benefits. One of the most commonly applied methods for studying both positive and negative attitudes toward AI is the use of questionnaires (Krägeloh et al., 2019). They have been used, among other things, in the context of education (Alves-Oliveira et al., 2015) and pedagogy (Peca et al., 2016). Based on the current measurements of attitudes toward AI, we conducted a critical review of existing questionnaires for assessing attitudes toward AI, before introducing the new one.

Attitudes toward AI are multidimensional (Park and Woo, 2022). Key psychological measurements include trust in AI (Liehner et al., 2023), intention to use (Chakim et al., 2023), acceptance or fear (Sindermann et al., 2022), and perceived usefulness (Maberah et al., 2025). In general population studies, widely used tools include the *General Attitudes toward Artificial Intelligence Scale* (GAAIS) (Schepman and Rodway, 2020) and the *Attitude Towards Artificial Intelligence* (ATAI) scale (Sindermann et al., 2021). Both instruments measure positive dimensions, such as usefulness and excitement, as well as negative ones, including fear and distrust. These tools demonstrate high internal reliability (Cronbach's α > 0.80) and confirmed factorial validity. However, neither has been translated into Polish, the language in which our research was conducted, nor validated for cultural sensitivity.

It is also worth mentioning the *Theory of Trust and Acceptance of Artificial Intelligence Technology* (TrAAIT), constructed to assess trust and acceptance of AI among clinicians (Stevens and Stetson, 2023), and the *Questionnaire of AI Use Motives* (QAIUM), developed for assessing students' motivations for using AI in educational settings (Yurt and Kasarci, 2024). Some instruments used in recent research (Na et al., 2022; Wang et al., 2023; Sohn and Kwon, 2022) are based on theoretical models such as the *Technology Acceptance Model* (TAM) (Davis et al., 1989) and the *Unified Theory of Acceptance and Use of Technology* (UTAUT) (Venkatesh et al., 2016). To some extent, this also applies to the questionnaire we propose below. Tools such as ATTARI-12 (Stein et al., 2024; Juranek, 2025) focus on general attitudes toward AI as a single construct, while others, such as the aforementioned GAAIS and ATAI, distinguish between positive and negative components, albeit in a broad and nonspecific manner. Items in the ATAI scale are primarily based on very general statements regarding unspecified forms of artificial intelligence; similarly, items in the GAAIS scale are too general to evaluate a specific AI type, such as the generative artificial intelligence assessed by respondents in our study. One of the most recent questionnaires was proposed by Park and colleagues (2024). They constructed a multidimensional scale called the Attitudes Towards AI Application at Work to measure workers' psychological attitudes toward AI in the workplace, covering cognitive and affective components such as perceived humanlikeness, adaptability, quality of AI, anxiety about using AI, job insecurity, and personal utility. Although this scale is the closest to our research agenda, its long list of items and lack of a validated and Polish version directed us towards proposing a new scale (described later in the text).

Not all of the tools discussed above account for social concerns related to AI, which seems particularly important, including those regarding automation of labor, algorithmic bias, or the ethics of AI-related decision-making i.e., decisions made by AI, decisions regarding AI use, and decisions made by those who design AI (Frey and Osborne, 2017; Buolamwini and Gebru, 2018). Such concerns can significantly shape attitudes toward AI, yet current scales tend to neglect or only capture them in a fragmented manner (Cave et al., 2019). Moreover, many tools, such as GAAIS, ATAI, or ATTARI-12, primarily focus on traditional systems for decision automation, prediction, or pattern recognition, rather than the currently popular generative artificial intelligence (Hussain et al., 2025), which is the focus of our study.

As GenAI tools, capable of producing text, visual, and audio content, become increasingly prominent, it is evident that public attitudes toward this new class of technology may differ qualitatively from those previously measured and require new measurement tools. Given the anticipated integration of GenAI agents into everyday professional tasks, communication, and knowledge management raises questions about the moral implications (Zlateva et al., 2024). These issues broadly fall outside the scope of existing generalized measures of AI attitudes, which further motivates the development of the new AGAWA scale presented below, taking into account cooperation with GenAI at work and considering the moral aspect.

The arguments outlined above underscore the need for a tool tailored to the current landscape of GenAI development and implementation in the workplace, which will consider both the pragmatic potential of this technology and the concerns it raises. Few existing scales, such as ATTARI-12 (Stein et al., 2024), though designed to measure general attitudes toward AI as a one-dimensional construct, can be meaningfully adapted to the GenAI context. However, they do not explicitly consider the future dimension of readiness to use generative systems as collaborators, nor the moral concerns associated with their application. Furthermore, given the growing popularity of GenAI in workplace environments (Microsoft and LinkedIn, 2024; McKinsey, 2024), there is a lack of questionnaires that assess both evaluation and readiness to adopt this type of technology.

## Model and Hypotheses

The aim of this paper was to identify barriers to implementing GenAI in the workplace, treating it as a work collaborator, and the readiness to view artificial intelligence as a valuable collaborator. In particular, we consider the three following factors possibly influencing a negative attitude toward AI: 1) concerns about interaction with GenAI, 2) anxiety related to its human-like characteristics, and 3) a belief in the uniqueness of human nature or even superiority over GenAI.

Based on the above literature review, we can make the following assumptions about the relationship between attitudes toward GenAI and these factors. The inherently interactive nature of collaboration, including those with GenAI, suggests a strong link between openness to interaction and the quality of cooperation, which appears justified (Baek and Kim, 2023). In contrast, the concerns about the impact of AI on society should correlate negatively with willingness to cooperate (Kelley et al., 2021; Gerlich, 2023). It refers less directly to human-AI interaction and more to general acceptance or rejection of GenAI in social life. Apart from recognizing GenAI as part of the work environment, being open to interaction is also crucial for its acceptance as a collaborator. The negative correlation between attitude toward GenAI and the perceived uniqueness of humans relative to artificial intelligence was also expected, as it may negatively influence the course of human-AI cooperation (Li et al., 2024; Zhang and Gosline, 2023).

We incorporate the impact of these factors in the model presented in Fig. 1. We hypothesize that all of the above factors have a negative impact on attitudes toward GenAI, as defined in Section "Construct of Attitudes toward GenAI", which will influence the willingness to use this technology, measured by frequency of its use (Rosen et al., 2013; Fakhri et al., 2024).

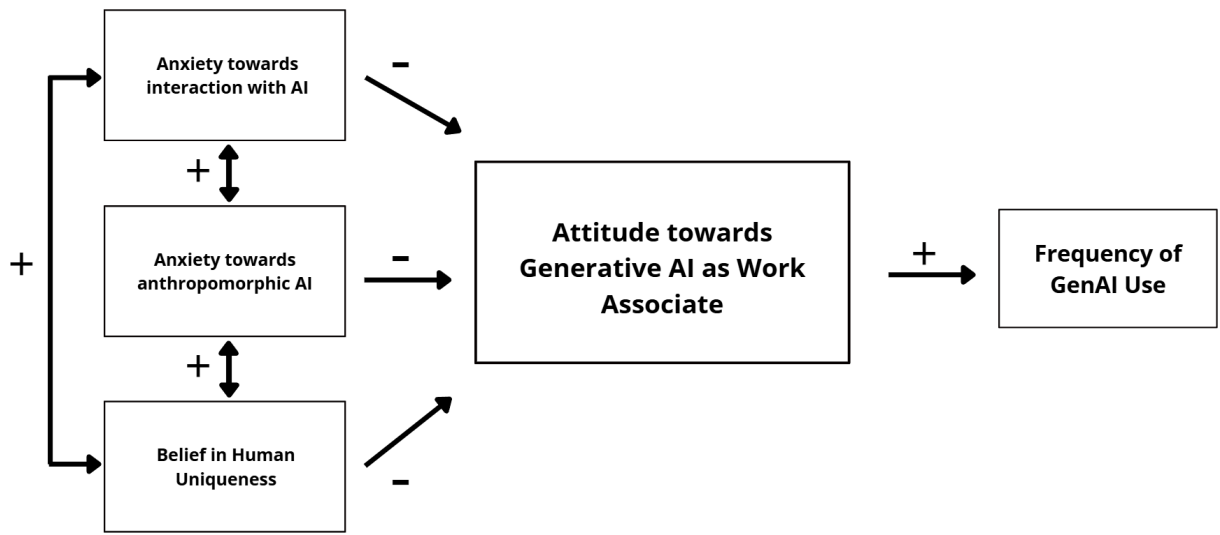

Figure 1. Model of relationships tested in this research. Signs "+" and "-" represent expected positive and negative correlations.

In particular, we hypothesize that:
> H1. The level of anxiety about interacting with an artificial agent utilising GenAI negatively correlates with the attitude towards this agent.
> H2. The level of anxiety about an artificial agent utilizing GenAI exhibiting human characteristics correlates negatively with the attitude towards it.
> H3. The sense of human uniqueness towards an artificial agent utilizing GenAI correlates negatively with attitude towards this agent.
> H4. The factors listed above correlate positively with each other.
> H5. Finally, attitude toward GenAI positively correlates with the declared frequency of use of this technology.

To measure interactions with artificial agents, fears associated with agents that exhibit human characteristics, and the sense of human uniqueness, we will use modified scales described below.

# Method

## Participants and Recruitment

The study participants were students from many Polish universities. Information about the study was distributed prior to classes in the form of a slide displayed on the screen with a QR code linking to the online questionnaire, through flyers posted in student groups on Facebook and LinkedIn, and through emails sent by the university's mail system. The data collection took approximately one and a half months, from the beginning of February to mid-March 2025. A total of 316 responses were collected. We excluded individuals who were not students, as well as all respondents over the age of 31, which exceeds the average

(M = 21.6) by three standard deviations (SD = 3.44). It resulted in a final sample of 296 participants aged 18 to 30 (21.2 ± 1.88), including 174 women, 99 men, 9 non-binary individuals, and 14 participants who did not disclose their gender. The participants represented 24 higher education institutions and various academic disciplines, categorized according to the OECD standard (InCites, 2022): *1 Natural sciences* - 19, *2 Engineering and technology* - 38, *3 Medical and health sciences* - 9, *5 Social sciences* - 178 (mostly *5.01 Psychology and cognitive sciences* - 114 and *5.03 Education* - 47), *6 Humanities and the arts* - 49, and 3 students who did not provide this information. The data from the questionnaire completed by students has been published in a public repository (Sikorski et al., 2025).

## NARS and BHNU Scales Modifications

We measured the three above-mentioned factors influencing attitudes toward artificial agents supported by GenAI using adapted scales described in the "Introduction", originally designed to investigate attitudes toward robots. These are: BHNU, which measures human uniqueness in comparison to robots, NATIR, which measures fears of interacting with robots, and NARHT - the fear of robots with human characteristics. The latter two scales are part of the NARS questionnaire. These three scales meet two conditions that are important for this research: 1) the existence of a Polish version of the questionnaire, preferred validated ones, and 2) documented evidence of its adaptability to other types of artificial agents.

As mentioned, all three scales have been modified to describe interaction with, fears of, and superiority over artificial agents instead of robots. Therefore, in used instruments, we substituted the term "robot" with "artificial intelligence" (cf. Fitrianie et al., 2020; Marchesi et al., 2019). Additionally, in the case of the NARS scale, we removed the item "I would feel very nervous just standing in front of a robot". This decision was based on the semantic inconsistency of this statement after substituting the word "robot" with "artificial intelligence," which could have led to participants' misinterpretation and reduced the results' reliability. The NARS questionnaire has already been modified by removing unsuitable items, e.g., for adolescents (De Graaf, & Allouch, 2013; cf. Park and Kim, 2024). As a result of these modifications, we refer to the adapted Polish version of the NARS questionnaire as NARS-AI-PL. The original Polish version of the NARS questionnaire (Pochwatko et al., 2015) comprises two subscales: NATIR and NARHT. After modification, the NATIR subscale should be interpreted as emotional discomfort and fear of interaction with AI, and therefore, we refer to it as NATIR-AI-PL. The modified NARHT subscale should be interpreted as concerns about AI with human-like features such as emotions or the ability to make judgments, and will be referred to as NARHT-AI-PL. In turn, the original version of BHNU measures the sense of human uniqueness in interactions with robots, which, after modification, will transform into a sense of uniqueness in interactions with artificial agents and is referred to as BHNU-AI-PL.

We assessed the reliability of modified scales by calculating Cronbach's alpha for both subscales using the collected data. For the NARHT-AI-PL subscale, Cronbach's alpha is 0.73, for the NATIR-AI-PL - 0.83, and for the BHNU-AI-PL - 0.83, which confirms their internal consistency.

# Questionnaire to Measure Attitudes to Generative AI

The online questionnaire comprised several sections. The first one collected basic demographic data and information about participants' studies. The next included items from the scales described above. In a separate section, there were eight statements on attitudes toward and readiness to use GenAI, forming the basis of the proposed AGAWA scale. Participants rated them using a seven-point Likert scale (1 - "Strongly disagree," 7 - "Strongly agree"; see Table 1).

**Table 1.** Statements regarding attitudes toward GenAI. Original Polish versions with English translations.

|    | Polish version | English version |
|----|----------------|-----------------|
| S1 | Firmy korzystające z pomocy generatywnej sztucznej inteligencji, w niedalekiej przyszłości zdobędą przewagę nad firmami nie korzystającymi z tej technologii. | Companies that utilize generative artificial intelligence will gain a competitive advantage over those that do not in the near future. |
| S2 | Firmy powinny korzystać z pomocy generatywnej sztucznej inteligencji w jak największym stopniu (oczywiście tylko w ramach legalnych działań), aby osiągnąć jak największą wydajność. | Companies should use generative artificial intelligence as extensively as possible - of course, only within legal bounds - to achieve maximum efficiency. |
| S3 | Rozwój generatywnej sztucznej inteligencji nie zagraża mojej przyszłej pracy. | The development of generative artificial intelligence does not pose a threat to my future job. |
| S4 | Generatywna sztuczna inteligencja okaże się wsparciem w mojej pracy zawodowej. | Generative artificial intelligence will prove to be a support in my professional work. |
| S5 | Nie obawiam się, że generatywna sztuczna inteligencja może niebawem w pełni zastąpić pracowników umysłowych. | I am not worried that generative artificial intelligence might soon replace knowledge workers entirely. |
| S6 | Uważam, że generatywna sztuczna inteligencja nie przewyższy możliwościami intelektualnymi ludzi, jeszcze za mojego życia. | I believe that generative artificial intelligence will not surpass human intellectual abilities within my lifetime. |
| S7 | Używanie generatywnej sztucznej inteligencji jako pomocy w pracy zawodowej nie budzi mojego moralnego oporu. | Using generative artificial intelligence as an assistance at professional work does not raise my moral objection. |
| S8 | Potrafiłbym/potrafiłabym przedstawić komuś sposób działania generatywnej sztucznej inteligencji. | I would be able to explain to someone how generative artificial intelligence works. |

The statements shown in Table 1 are a revised and expanded version of items previously used in our prior study on attitudes toward GenAI in the GameDev industry (Sikorski et al., 2025). In the current version, to avoid potential confusion, the items were reformulated so that higher values on the Likert scale consistently reflect more positive attitudes toward GenAI. Also, statements S3 and S4 were created as refinements of one earlier item. Additionally, we introduced items S5 through S8 to better capture respondents' views on the intellectual capabilities of GenAI, their moral stance toward it, and their subjective knowledge about the technology.

When compiling the survey statements, we followed the models of the two most recognised scales measuring general attitudes towards technology acceptance: TAM (Davis at al., 1989) and UTAUT (Venkatesh et al., 2003). Consequently, statements S1 and S2 refer to perceived usefulness from TAM and performance expectancy from UTAUT, with the latter also referring to social influence from UTAUT through its reference to the concept of duty. Statements S3 and S7 also refer to the social aspect. Statement S4 combines perceived usefulness from TAM, as well as the performance expectancy, effort expectancy, and facilitating conditions from UTAUT. Statements S5 and S6 also refer to the social aspect, specifically the fear of losing one's job and the fear of losing one's unique position as a human being, both strongly associated with the AI revolution and are not directly addressed in the TAM and UTAUT questions. The entire set of statements S1-S8, which measures attitudes towards AI and intention to use it, especially S2 and S4, refers to the attitude towards use (from TAM) and behavioural intention to use (from TAM and UTAUT) scales, which are implicitly included in the statements.

We measured the willingness to use GenAI by declared frequency of its use, namely by asking, 'How often do you use tools based on generative artificial intelligence?' ("Jak często używasz narzędzi opartych o generatywną sztuczną inteligencję?" in Polish), and provided respondents with the following answer options: 'Never', 'Once or twice a year', 'Once or twice a month', 'Once or twice a week', 'Almost every day or every day', and we assigned them values from 1 to 5 during the analysis. This reflects "actual system usage" according to TAM and "user behavior" according to UTAUT.

# The Construction of the AGAWA Scale

The collected data, particularly the responses to eight items regarding attitudes toward GenAI in the workplace, were analyzed in the following stages to prepare a scale measuring attitudes towards GenAI:
1. principal component analysis (PCA) to extract a subset of items forming a coherent scale describing attitudes toward GenAI as a potential workplace associate (AGAWA);
2. internal consistency analysis using Cronbach's alpha;
3. expert evaluation of AGAWA scale statements.

We then proceeded to verify the model described above, which relates attitudes toward GenAI to the examined factors: fears measured by NATIR-AI-PL and NARHT-AI-PL, as well as the sense of uniqueness measured by BHNU-AI-PL. We calculated the correlation of the results from these scales. Our goal was to create a short scale that integrates both personal believes about using AI in the workplace, trust towards AI as well as moral approaches in using generative AI in one structurally coherent construct.

## The Identifying Scale Statements and Checking Its Reliability

Principal component analysis (PCA) revealed a first component clustering items S1, S2, S4, and S7 with loadings of similar values and greater than 0.75 (0.770, 0.796, 0.776, and 0.769, respectively), indicating a strong common factor. These items were therefore combined into the AGAWA scale, calculated as the arithmetic mean of responses to the four items, assuming equal weights. It has been confirmed by Confirmatory Factor Analysis (CFA): $\chi^2$ = 38.1, $p$ = 0.004; CFI = 0.959 > 0.95, RMSEA = 0.0614 < 0.08, and SRMR = 0.0543 < 0.08.

The remaining items (S3, S5, S6, S8) had lower loadings on this component or loaded more strongly on other components, indicating that they represent different dimensions and were therefore excluded from the AGAWA scale. PCA analysis also indicated consistency between statements S3, S5, and S6 (loadings 0.634, 0.732, and 0.704, respectively), but with very low internal consistency (Cronbach's α value of less than 0.5).

Next, we assessed the scale's internal consistency, obtained in the previous step, by calculating Cronbach's α for the selected four items. It was equal to 0.804. Additionally, we examined whether removing any item could enhance internal consistency. In all cases, the α value decreased (the highest being 0.787 after removing the first item).

## Expert Evaluation of AGAWA Validity

A validation involved a survey conducted among experts, who assessed whether the four items comprising the newly developed AGAWA scale were conceptually related to attitudes toward generative artificial intelligence as a potential workplace collaborator. To this end, four additional items were included alongside items S1, S2, S4, and S7. These supplementary items also referred to GenAI but in slightly different contexts: A1 - "Generative AI is an effective tool for acquiring knowledge about the world"; A2 - "Using generative AI does not raise concerns in me about the potential for technological addiction"; A3 - "The development of generative AI will bring humanity more benefits than harm"; A4 - "The development of generative AI will enable almost anyone to become an artist". Respondents could answer each item with one of three options: "No", "I don't know", or "Yes", indicating the relevance of each statement to the studied construct. The survey included nine participants. A relevance rating of 70% or higher was adopted as the threshold for validity. All four AGAWA scale items met this criterion (S1 - 78%, S2 - 78%, S4 - 100%, S7 - 89%). In contrast, none of the additional items reached the 70% threshold (A1 - 56%, A2 - 56%, A3 - 22%, A4 - 11%).

The statistical properties of the AGAWA scale as calculated from the data collected in the present study (n = 296) are as follows. The mean value is equal to M = 4.30 (for a range from 1 to 7) and the standard deviation - SD = 1.41 (see Fig. 1). The median was 4.5. The distribution of values did not conform to a normal distribution (Shapiro-Wilk test, p = 0.001); skewness was -0.193, and kurtosis was -0.552. Note that both the mean and the median were above the middle of the scale, which was equal to 4.0.

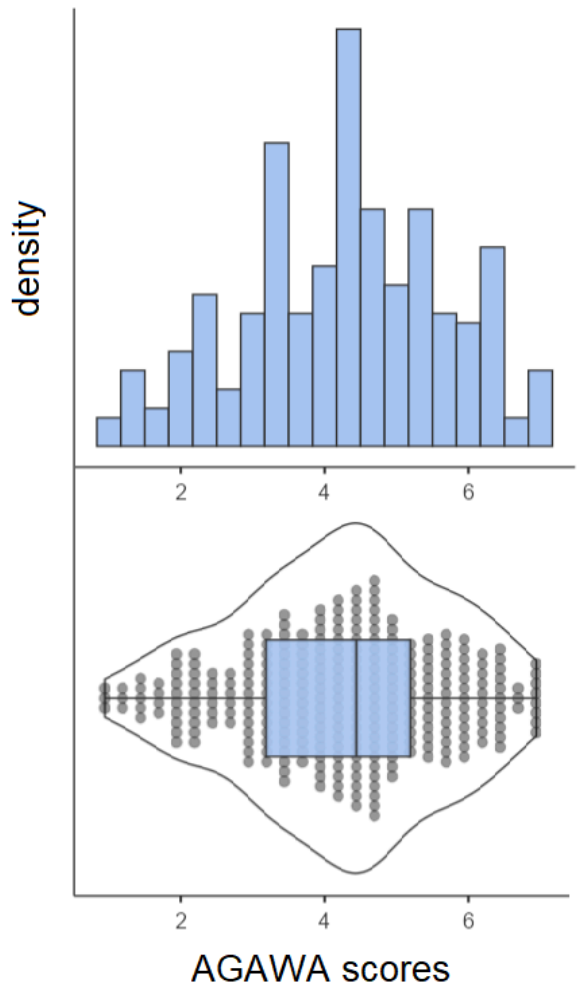

**Fig. 1.** Distribution of AGAWA Scale Scores

# Results

We examined the correlations between the AGAWA scale results and the BHNU-AI-PL scale, as well as two NARS-AI-PL subscales. It is important to note again that these scales indicate negative attitudes toward AI; therefore, we expected negative correlations with the proposed AGAWA scale, which reflects approval of GenAI. The results are presented in Table 2 and Fig. 3. As expected, the attitude toward AI as a coworker negatively correlates with anxiety of interaction with AI (r = -0.562, p < .001), the AI with humanlike characteristics (r = -0.435, p < 0.001) and a sense of human uniqueness (r = -0.225, p < .001) are negative and statistically significant, which confirms hypotheses H1, H2, and H3. At the same time, the above factors correlate positively with each other (see Table 2), which supports hypothesis H4. Finally, a positive attitude towards GenAI is correlated with the frequency of using this tool (r = 0.568, p < .001), which confirms hypothesis H5.

**Table 2.** Statistical correlations (*r*) with statistical significance (*p*) between AGAWA and the compared items

| Compared groups | AGAWA | | NATIR-AI-PL | | NARHT-AI-PL | |
|---|---|---|---|---|---|---|
| | r | p | r | p | r | p |
| NATIR-AI-PL | -0.562*** | <.001 | | | | |
| NARHT-AI-PL | -0.435*** | <.001 | 0.661*** | <.001 | | |
| BHNU-AI-PL | -0.225*** | <.001 | 0.495*** | <.001 | 0.500** | <.001 |
| Frequency of GenAI Use | -0.568*** | <.001 | | | | |

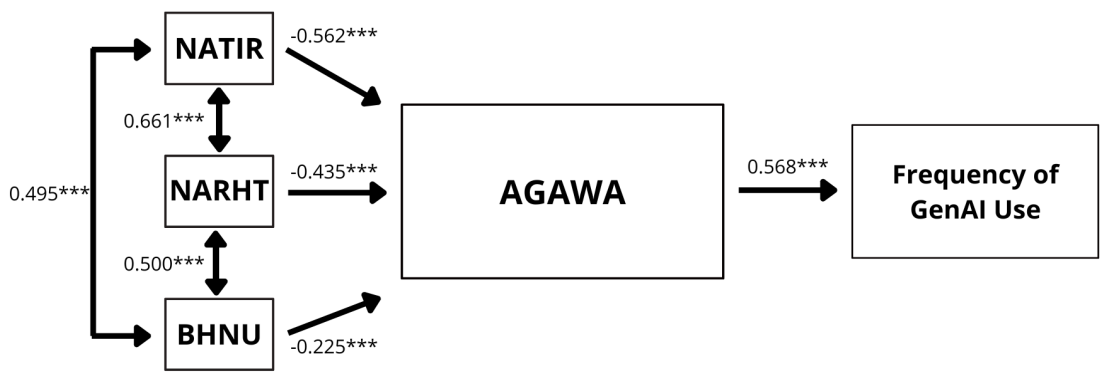

Fig. 3. Relationships between model elements

## Discussion

The consistency of the results obtained using the questionnaires, including three scales relating to the concerns that make up the factors under study and a new scale measuring attitudes toward GenAI in the workplace, suggests the existence of a consistent attitude toward GenAI. It determines the propensity, consent, and willingness to treat GenAI as a co-worker. Strong mutual correlations between the results of the NATIR-AI-PL, NARHT-AI-PL, and BHNU-AI-PL scales (cf. Giger et al., 2025; Łupkowski and Gierszewska, 2019) indicate that this attitude may be significantly shaped by the fears felt by the respondents. Respondents who express concerns about interacting with a GenAI agent are more likely to evaluate its potential human characteristics negatively and, at the same time, believe more strongly in the uniqueness of humans as beings that are irreplaceable by GenAI. Therefore, we can speak of the existence of a coherent set of beliefs and attitudes that reinforce each other, thus hindering the development of a positive attitude towards cooperation with GenAI.

As mentioned above, our results confirm that negative attitudes toward GenAI are strongly associated with fear of interacting with AI and concern about its anthropomorphic features (in line with the assumptions of the NARS scale; Nomura et al., 2006; Pochwatko et al., 2015), which is consistent with previous research on aversion to artificial agents and avatars (Diana et al., 2022; Park & Kim, 2024). At the same time, as noted by Giger et al. (2017, 2025) and Łupkowski and Gierszewska (2019), a strong belief in human uniqueness in relation to machines is a significant obstacle to the acceptance of artificial intelligence as an equal collaborator. Also in our results, a statistically significant correlation between these factors is evident, though it is weak; therefore our results confirm that a sense of human uniqueness correlates negatively with the willingness to collaborate with GenAI, but this is not a strong relationship.

It is interesting to compare the strength of correlation between the three factors under consideration, which will allow us to assess their importance in shaping attitudes toward GenAI. The results obtained indicate that the emotional aspects of contact with GenAI are more critical to attitudes toward AI than the sense of human uniqueness.

The inherently interactive nature of collaboration, including those with GenAI, suggests a strong link between openness to interaction and the quality of cooperation (Baek and Kim, 2023). This strong relationship was confirmed in our study. In contrast, the concerns about the impact of AI on society should correlate negatively with willingness to cooperate (Kelley et al., 2021; Gerlich, 2023). And this is also confirmed in this study. It refers less directly to human-AI interaction and more to general acceptance or rejection of GenAI in social life. This is the possible explanation why the correlation between a positive attitude towards the use of GenAI in the workplace and fear of GenAI agents exhibiting human characteristics is weaker than that with fear of interaction with GenAI, and why recognizing GenAI as part of the work environment is vital for its acceptance as a collaborator; openness to interaction is more important.

The other examined factor is a tendency to perceive humans as distinct among other beings, including artificial ones. As mentioned in the "Introduction", other studies have shown a relationship between the belief in human uniqueness and negative attitudes toward robots (Giger et al., 2024; Złotowski et al., 2017). Our research confirms that this is also true for artificial agents based on GenAI. A possible explanation for this relationship is that individuals who strongly endorse such beliefs tend to perceive human-AI boundaries as sacrosanct, and any suggestion that AI might encroach upon these boundaries evokes unease or moral resistance. The psychological justification is rooted in distinctiveness threat and ontological boundary defense: individuals scoring high in human uniqueness perceive human cognitive and moral capacities as the final bastion of humanness. When AI demonstrates creativity, empathy, or autonomous reasoning, it symbolically infringes upon this domain, eliciting identity-protective responses such as distrust, rejection, or moral exclusion. Thus, the correlation reflects a motivated effort to preserve the uniqueness and moral superiority of the human category in the face of advancing artificial agents (Ferrari et al., 2016; Giger et al., 2024; Pochwatko et al., 2015; Złotowski et al., 2017).

Our findings, concerning negative attitudes toward AI with human-like characteristics and the willingness to interact with it, also seem relevant in the context of the ongoing discussion about the superiority of artificial intelligence over humans in such fundamental areas as intelligence (Korteling et al., 2021), task performance (Hemmer et al., 2023), and decision-making speed (Varma et al., 2023). We confirmed the negative influence of belief in the uniqueness of human nature on the willingness to use generative AI.

Finally, it is worth noting that, as in some previous studies suggesting a moderately positive or ambivalent attitude toward AI in the general population (Gillespie et al., 2021; Schepman and Rodway, 2020), we obtained results for the AGAWA scale that are close to the middle of the scale. Our average was shifted toward the positive, but only slightly. These results confirm mixed attitudes toward AI, with a slight predominance of optimism in groups with higher exposure to the technology.

# Conclusions

The results of this study directly indicate that reducing resistance to interacting with GenAI and addressing concerns about its anthropomorphic features should lead to an increased willingness within the students to collaborate with GenAI in the workplace and a higher adoption rate of this technology.

The above findings indicate that collaboration with GenAI in the professional environment cannot occur without readiness for interaction with this technology. Furthermore, to encourage future employees to collaborate with GenAI, that is, to acknowledge its usefulness and reduce moral resistance, it is essential first to prompt them to interact with an AI agent. A typical example is when employers require knowledge workers to begin working with GenAI to increase productivity. It is very important to minimize employee discomfort accompanying the implementation of new technology. An excellent first step is to introduce them to artificial agents, for example, through games that involve collaboration with GenAI or incorporate it into team problem-solving tasks. Shaping positive attitudes toward GenAI may also begin in preschool and school education, in a manner analogous to how basic computer skills are currently taught.

It is worth noting that the presence of GenAI and the required attitude toward collaboration with it extend beyond the workplace. This technology is already being used in banking, commerce, transportation, and other domains, which can be considered generalized forms of cooperation.

To achieve the research objective, we have developed a short AGAWA scale measuring attitudes toward GenAI as a collaborator. Our scale, consisting of only four items, offers a substantial advantage in terms of administration time, enabling the assessment of attitudes toward artificial intelligence in just about one minute. Full description of AGAWA with additional comments and list of items can be found in Appendix.

# Limitations and Future Work

The presented study was conducted with student participants from Poland. A natural next step is to expand the sample of participants to a broader cross-section of society, with a focus on knowledge workers, whose professional activities are already influenced by GenAI. Respondents from other countries should also be included to examine potential cultural differences in attitudes towards AI and the barriers to its implementation in the workplace. This will require validating the AGAWA scale in other language versions, including English.

# Declaration of generative AI and AI-assisted technologies in the writing process

During the preparation of this work, the authors used ChatGPT and Copilot in order to help with translation, grammar, and spelling correction. After using this service, the authors reviewed and edited the content as needed and take full responsibility for the content of the published article.

# References

1. Almaraz-López C., Almaraz-Menéndez F., López-Esteban C. (2023), Comparative Study of the Attitudes and Perceptions of University Students in Business Administration and Management and in Education toward Artificial Intelligence
2. Alves-Oliveira, P., Ribeiro, T., Petisca, S., Di Tullio, E., Melo, F. S., and Paiva, A. (2015). An empathic robotic tutor for school classrooms: Considering expectation and satisfaction of children as end-users. In Social Robotics: 7th International Conference, ICSR 2015, Paris, France, October 26-30, 2015, Proceedings 7 (pp. 21-30). Springer International Publishing.
3. Baek, T. H., & Kim, M. (2023). Is ChatGPT scary good? How user motivations affect creepiness and trust in generative artificial intelligence. Telematics and Informatics, 83, 102030.
4. Bergdahl J., Latikka R., Celuch M., Savolainen I., Soares Mantere E., Savela N., Oksanen A., (2023), Self-determination and attitudes toward artificial intelligence: Cross-national and longitudinal perspectives, Telematics and Informatics
5. Boucher, J., Smith, G., Telliel, Y. D., (2024), Is Resistance Futile?: Early Career Game Developers, Generative AI, and Ethical Skepticism, Conference on Human Factors in Computing Systems - Proceedings, Article No.: 173, Pages 1 - 13, DOI:10.1145/3613904.3641889
6. Brynjolfsson, E., Li, D., & Raymond, L. (2025). Generative AI at work. The Quarterly Journal of Economics, 140(2), 889-942.
7. Buolamwini, J., and Gebru, T. (2018, January). Gender shades: Intersectional accuracy disparities in commercial gender classification. In Conference on fairness, accountability and transparency (pp. 77-91). PMLR.
8. Cardon P. W., Getchell K., Carradini S., Fleischmann C., Stapp J., (2023), Generative AI in the Workplace: Employee Perspectives of ChatGPT Benefits and Organizational Policies, https://osf.io/preprints/socarxiv/b3ezy
9. Cave, S., Coughlan, K., and Dihal, K. (2019, January). " Scary Robots" Examining Public Responses to AI. In Proceedings of the 2019 AAAI/ACM Conference on AI, Ethics, and Society (pp. 331-337).
10. Chakim, M. H. R., Kho, A., Santoso, N. P. L., and Agustian, H. (2023). Quality factors of intention to use in artificial intelligence-based aiku applications. ADI Journal on Recent Innovation, 5(1), 72-85.
11. Dahmash, A. B., Alabdulkareem, M., Alfutais, A., Kamel, A. M., Alkholaiwi, F., Alshehri, S., Zahrani, Y. A., Almoaiqel, M., (2020), Artificial intelligence in radiology: does it impact medical students preference for radiology as their future career?,

BJR|Open, Volume 2, Issue 1, 1 November 2020, 20200037, https://doi.org/10.1259/bjro.20200037
12. Davis, F. D., Bagozzi, R. P., & Warshaw, P. R. (1989). Technology acceptance model. J Manag Sci, 35(8), 982-1003.
13. De Graaf, M. M., & Allouch, S. B. (2013, August). The relation between people's attitude and anxiety towards robots in human-robot interaction. In 2013 IEEE RO-MAN (pp. 632-637). IEEE.
14. Diana, F., Kawahara, M., Saccardi, I. et al. A Cross-Cultural Comparison on Implicit and Explicit Attitudes Towards Artificial Agents. Int J of Soc Robotics 15, 1439-1455 (2023). https://doi.org/10.1007/s12369-022-00917-7
15. Fakhri, M. M., Ahmar, A. S., Isma, A., Rosidah, R., & Fadhilatunisa, D. (2024). Exploring generative AI tools frequency: Impacts on attitude, satisfaction, and competency in achieving higher education learning goals. EduLine: Journal of Education and Learning Innovation, 4(1), 196-208.
16. Ferrari, F., Paladino, M. P., & Jetten, J. (2016). Blurring human-machine distinctions: Anthropomorphic appearance in social robots as a threat to human distinctiveness. International Journal of Social Robotics, 8(2), 287-302.
17. Frey, C. B., and Osborne, M. A. (2017). The future of employment: How susceptible are jobs to computerisation?. Technological forecasting and social change, 114, 254-280.
18. Fitrianie, S., Bruijnes, M., Richards, D., Bönsch, A., & Brinkman, W. P. (2020, October). The 19 unifying questionnaire constructs of artificial social agents: An iva community analysis. In Proceedings of the 20th ACM International Conference on Intelligent Virtual Agents (pp. 1-8).
19. GDC, gamedeveloper.com (2024), 2024 State of the Game Industry, https://images.reg.techweb.com/Web/UBMTechweb/%7B4fe03be1-d6b1-4f91-95f4-b4bdea9e739e%7D_GDC24-SOTI-Report_Final.pdf
20. Gerlich, M. (2023). Perceptions and acceptance of artificial intelligence: A multi-dimensional study. Social Sciences, 12(9), 502.
21. Giger, J. C., Moura, D., Almeida, N., and Piçarra, N. (2017, May). Attitudes towards social robots: The role of gender, belief in human nature uniqueness, religiousness and interest in science fiction. In Proceedings of II International Congress on Interdisciplinarity in Social and Human Sciences (Vol. 11, p. 509).
22. Giger, J. C., Piçarra, N., Pochwatko, G., Almeida, N., & Almeida, A. S. (2025). Intention to Work with Social Robots: The Role of Perceived Robot Use Self-Efficacy, Attitudes Towards Robots, and Beliefs in Human Nature Uniqueness. Multimodal Technologies and Interaction, 9(2), 9.
23. Giger, J. C., Piçarra, N., Pochwatko, G., Almeida, N., Almeida, A. S., and Costa, N. (2024). Development of the beliefs in human nature uniqueness scale and its associations with perception of social robots. Human Behavior and Emerging Technologies, 2024(1), 5569587.
24. Gillespie, N., Lockey, S., Curtis, C., (2021), Trust in artificial Intelligence: a five country study, The University of Queensland, https://espace.library.uq.edu.au/view/UQ:e34bfa3
25. Gmyrek, P., Berg, J., Kamiński, K., Konopczyński, F., Ładna, A., Nafradi, B., Rosłaniec, K, Troszyński, M. (2025). Generative AI and jobs: A refined global index of occupational exposure (No. 140). ILO Working Paper.

https://www.ilo.org/publications/generative-ai-and-jobs-refined-global-index-occupational-exposure
26. Granulo, A., Fuchs, C., & Puntoni, S. (2019). Psychological reactions to human versus robotic job replacement. Nature human behaviour, 3(10), 1062-1069.
27. Hemmer, P., Westphal, M., Schemmer, M., Vetter, S., Vössing, M., & Satzger, G. (2023, March). Human-AI collaboration: the effect of AI delegation on human task performance and task satisfaction. In Proceedings of the 28th International Conference on Intelligent User Interfaces (pp. 453-463).
28. Hopelab, Common Sense Media, Harvard Graduate School of Education, Center for Digital Thriving, (2024), Teen and Young Adult Perspectives on Generative AI, https://pz.harvard.edu/resources/teen-and-young-adult-perspectives-generative-ai
29. Hussain, A., Ali, S., Farwa, U. E., Mozumder, M. A. I., and Kim, H. C. (2025, February). Foundation Models: From Current Developments, Challenges, and Risks to Future Opportunities. In 2025 27th International Conference on Advanced Communications Technology (ICACT) (pp. 51-58). IEEE.
30. InCites, (2022), OECD schema to Web of Science Category Mapping 2022, https://incites.zendesk.com/hc/en-gb/articles/22516984338321-OECD-Category-Schema
31. International Labour Organization, (2024), World Employment and Social Outlook, Trends 2024, https://www.ilo.org/publications/flagship-reports/world-employment-and-social-outlook-trends-2024
32. Kelley, P. G., Yang, Y., Heldreth, C., Moessner, C., Sedley, A., Kramm, A., ... & Woodruff, A. (2021, July). Exciting, useful, worrying, futuristic: Public perception of artificial intelligence in 8 countries. In Proceedings of the 2021 AAAI/ACM Conference on AI, Ethics, and Society (pp. 627-637).
33. Juranek, J. F. (2025). Measurement of General Attitude Toward AI and its Relationship with AI Anxiety, Attitude Toward Robots and Belief in Human Nature Uniqueness: Polish Adaptation of the ATTARI-12 Questionnaire.
34. Karamthulla, M. J., Tadimarri, A., Tillu, R., and Muthusubramanian, M. (2024). Navigating the future: AI-driven project management in the digital era. International Journal for Multidisciplinary Research, 6(2), 1-11.
35. Kaya, F., Aydin, F., Schepman, A., Rodway, P., Yetişensoy, O., Demir Kaya, M., (2024), The Roles of Personality Traits, AI Anxiety, and Demographic Factors in Attitudes toward Artificial Intelligence, International Journal of Human-Computer Interaction.
36. Kodani, N., Uchida, T., Kameo, N., Ban, M., Sakai, K., Funayama, T., Minato, T., Kikuchi, A., and Ishiguro, H. (2024). Effects of Gestures and Negative Attitudes on Impressions of Lecturing Robots. In Companion of the 2024 ACM/IEEE International Conference on Human-Robot Interaction (pp. 628-631). HRI '24: ACM/IEEE International Conference on Human-Robot Interaction. ACM. https://doi.org/10.1145/3610978.3640612
37. Korteling, J. E., van de Boer-Visschedijk, G. C., Blankendaal, R. A., Boonekamp, R. C., & Eikelboom, A. R. (2021). Human-versus artificial intelligence. Frontiers in artificial intelligence, 4, 622364.
38. Krägeloh, C. U., Bharatharaj, J., Sasthan Kutty, S. K., Nirmala, P. R., and Huang, L. (2019). Questionnaires to measure acceptability of social robots: a critical review. Robotics, 8(4), 88.

39. Latikka, R., Bergdahl, J., Savela, N., & Oksanen, A. (2023). AI as an artist? A two-wave survey study on attitudes toward using artificial intelligence in art. Poetics, 101, 101839.
40. Law, M., & Varanasi, R. A. (2025, May). Generative AI and Changing Work: Systematic Review of Practitioner-Led Work Transformations Through the Lens of Job Crafting. In International Conference on Human-Computer Interaction (pp. 131-152). Cham: Springer Nature Switzerland.
41. Liang, Y., Lee, S. A., (2017), Fear of Autonomous Robots and Artificial Intelligence: Evidence from National Representative Data with Probability Sampling, International Journal of Social Robotics.
42. Li, J., Cao, H., Lin, L., Hou, Y., Zhu, R., & El Ali, A. (2024, May). User experience design professionals' perceptions of generative artificial intelligence. In Proceedings of the 2024 CHI conference on human factors in computing systems (pp. 1-18).
43. Liehner, G. L., Biermann, H., Hick, A., Brauner, P., and Ziefle, M. (2023). Perceptions, attitudes and trust towards artificial intelligence—An assessment of the public opinion. Artificial Intelligence and Social Computing, 72, 32-41.
44. Lovato, J., Zimmerman, J. W., Smith, I., Dodds, P., Karson, J. L., (2024), Foregrounding Artist Opinions: A Survey Study on Transparency, Ownership, and Fairness in AI Generative Art, Proceedings of the AAAI/ACM Conference on AI, Ethics, and Society (pages 905-916).
45. Lupkowski, P., & Gierszewska, M. (2019). Attitude towards humanoid robots and the uncanny valley hypothesis. Foundations of Computing and Decision Sciences, 44(1), 101-119.
46. Maberah, S., Kan'an, A., El-Sayed, N., Alahmari, M., Abdelmabood, M., Kholif, M., ... and Alshehri, E. (2025). Students' Attitudes and Perceived Usefulness of Artificial Intelligence (AI) Tools in Physical Education. International Journal of Information and Education Technology, 15(4).
47. Marchesi, S., Ghiglino, D., Ciardo, F., Perez-Osorio, J., Baykara, E., & Wykowska, A. (2019). Do we adopt the intentional stance toward humanoid robots?. Frontiers in psychology, 10, 450.
48. McKinsey (2024), The state of AI in early 2024: Gen AI adoption spikes and starts to generate value, https://www.mckinsey.com/capabilities/quantumblack/our-insights/the-state-of-ai
49. Microsoft, LinkedIn, (2024). 2024 Work Trend Index Annual Report from Microsoft and LinkedIn, https://www.microsoft.com/en-us/worklab/work-trend-index/ai-at-work-is-here-now-comes-the-hard-part#:~:text=75%25%20of%20knowledge%20workers%20use,work%20more%20(83%25).
50. Milicevic, N., Kalas, B., Djokic, N., Malcic, B., and Djokic, I. (2024). Students' Intention toward Artificial Intelligence in the Context of Digital Transformation. *Sustainability*, *16*(9), 3554.
51. Na, S., Heo, S., Han, S., Shin, Y., and Roh, Y. (2022). Acceptance model of artificial intelligence (AI)-based technologies in construction firms: Applying the Technology Acceptance Model (TAM) in combination with the Technology-Organisation-Environment (TOE) framework. Buildings, 12(2), 90.
52. Nomura, T., Suzuki, T., Kanda, T., and Kato, K. (2006). Measurement of negative attitudes toward robots. Interaction Studies. Social Behaviour and Communication in Biological and Artificial Systems, 7(3), 437-454.

53. Park, C. J., Yi, P. H., Siegel, E. L., (2021), Medical Student Perspectives on the Impact of Artificial Intelligence on the Practice of Medicine, Current Problems in Diagnostic Radiology 50 (2021) 614619 DOI:10.1067/J.CPRADIOL.2020.06.011
54. Park, J., and Woo, S. E. (2022). Who likes artificial intelligence? Personality predictors of attitudes toward artificial intelligence. The Journal of Psychology, 156(1), 68-94.
55. Park, J., Woo, S. E., and Kim, J. (2024). Attitudes towards artificial intelligence at work: Scale development and validation. In Journal of Occupational and Organizational Psychology (Vol. 97, Issue 3, pp. 920-951). Wiley. https://doi.org/10.1111/joop.12502
56. Peca, A., Coeckelbergh, M., Simut, R., Costescu, C., Pintea, S., David, D., and Vanderborght, B. (2016). Robot Enhanced Therapy for Children with Autism Disorders. IEEE TEchnology and SocIETy MagazInE, 1932(4529/16).
57. Piçarra N., Giger J.-C., Pochwatko G. , Gonçalves G. , "Validation of the Portuguese version of the Negative Attitudes towards Robots Scale", Revue européenne de psychologie appliquée, vol. 65, 2015, 93-104. DO: 10.1016/j.erap.2014.11.002.1162-9088.
58. Pochwatko, G., Giger, J. C., Różańska-Walczuk, M., Świdrak, J., Kukiełka, K., Możaryn, J., and Piçarra, N. (2015). Polish version of the negative attitude toward robots scale (NARS-PL). Journal of Automation Mobile Robotics and Intelligent Systems, 9.
59. Ratajczyk, D., Dakowski, J., and Łupkowski, P. (2024). The importance of beliefs in human nature uniqueness for uncanny valley in virtual reality and on-screen. International journal of human-computer interaction, 40(12), 3081-3091.
60. Rosen, L. D., Whaling, K., Carrier, L. M., Cheever, N. A., & Rokkum, J. (2013). The media and technology usage and attitudes scale: An empirical investigation. Computers in human behavior, 29(6), 2501-2511.
61. Schepman, A., and Rodway, P. (2020). Initial validation of the general attitudes towards Artificial Intelligence Scale. Computers in human behavior reports, 1, 100014.
62. SignalFire, (2025), The SignalFire State of Tech Talent Report - 2025, https://www.signalfire.com/blog/signalfire-state-of-talent-report-2025
63. Sikorski, Ł., Matulewski, J., Czerwonka, M. (2025) On the Attitudes of GameDev Industry Artists towards GenAI. Preliminary Results. SIGMIS-CPR '25: Proceedings of the 2025 Computers and People Research Conference. Association for Computing Machinery, New York, NY, USA, Article 2, 1–6 (2025). doi:10.1145/3716489.3728432
64. Sikorski, Łukasz; Matulewski, Jacek; Łukasik, Albert (2025), "AGAWA Dataset", Mendeley Data, V2, doi: 10.17632/9zz4bb5rbs.2
65. Salari, N., Beiromvand, M., Hosseinian-Far, A., Habibi, J., Babajani, F., & Mohammadi, M. (2025). Impacts of generative artificial intelligence on the future of labor market: A systematic review. Computers in Human Behavior Reports, 100652.
66. Schwab, K. (2016). The Fourth Industrial Revolution. World Economic Forum. Cologny/ Geneva Switzerland : World Economic Forum REF: 231215, 1-172.
67. Sindermann, C., Sha, P., Zhou, M., Wernicke, J., Schmitt, H. S., Li, M., ... and Montag, C. (2021). Assessing the attitude towards artificial intelligence: Introduction of a short measure in German, Chinese, and English language. KI-Künstliche intelligenz, 35(1), 109-118.

68. Sindermann, C., Yang, H., Elhai, J. D., Yang, S., Quan, L., Li, M., and Montag, C. (2022). Acceptance and Fear of Artificial Intelligence: associations with personality in a German and a Chinese sample. Discover Psychology, 2(1), 8.
69. Sohn, K., and Kwon, O. (2020). Technology acceptance theories and factors influencing artificial Intelligence-based intelligent products. Telematics and Informatics, 47, 101324.
70. Stack Overflow, (2024), 2024 Developer Survey. AI, https://survey.stackoverflow.co/2024/ai/
71. Stein, J.-P., Liebold, B., and Ohler, P. (2019). Stay back, clever thing! Linking situational control and human uniqueness concerns to the aversion against autonomous technology. In Computers in Human Behavior (Vol. 95, pp. 73-82). Elsevier BV. https://doi.org/10.1016/j.chb.2019.01.021
72. Stein, J. P., Messingschlager, T., Gnambs, T., Hutmacher, F., and Appel, M. (2024). Attitudes towards AI: measurement and associations with personality. Scientific Reports, 14(1), 2909.
73. Stevens, A. F., and Stetson, P. (2023). Theory of trust and acceptance of artificial intelligence technology (TrAAIT): An instrument to assess clinician trust and acceptance of artificial intelligence. Journal of biomedical informatics, 148, 104550.
74. Syzygy, (2017), Sex, lies and A.I. How Americans feel about artificial intelligence What marketers need to know, https://daks2k3a4ib2z.cloudfront.net/59c269cb7333f20001b0e7c4/59db4483aaa78100013fa85a_Sex_lies_and_AI-SYZYGY-Digital_Insight_Report_2017_US.pdf
75. Varma, A., Dawkins, C., & Chaudhuri, K. (2023). Artificial intelligence and people management: A critical assessment through the ethical lens. Human Resource Management Review, 33(1), 100923.
76. Venkatesh, V., Thong, J. Y., and Xu, X. (2016). Unified theory of acceptance and use of technology: A synthesis and the road ahead. Journal of the association for Information Systems, 17(5), 328-376.
77. Venkatesh, V., Morris, M. G., Davis, G. B., & Davis, F. D. (2003). User acceptance of information technology: Toward a unified view. MIS quarterly, 425-478.
78. Vimpari, V., Kultima, A., Hämäläinen, P., and Guckelsberger, C. (2023). "An Adapt-or-Die Type of Situation": Perception, Adoption, and Use of Text-to-Image-Generation AI by Game Industry Professionals. Proceedings of the ACM on Human-Computer Interaction, 7(CHI PLAY), 131-164..
79. Wang, C., Ahmad, S. F., Ayassrah, A. Y. B. A., Awwad, E. M., Irshad, M., Ali, Y. A., ... and Han, H. (2023). An empirical evaluation of technology acceptance model for Artificial Intelligence in E-commerce. Heliyon, 9(8).
80. World Economic Forum (WEF), (2025), Future of Jobs Report 2025, https://www.weforum.org/publications/the-future-of-jobs-report-2025/
81. Yang, J., Kikuchi, H., Uegaki, T., Moriki, K., and Kikuchi, H. (2022). The Effect of the Repetitive Utterances Complexity on User's Desire to Continue Dialogue by a Chat-oriented Spoken Dialogue System. In Proceedings of the 10th International Conference on Human-Agent Interaction (pp. 51-56). HAI '22: International Conference on Human-Agent Interaction. ACM. https://doi.org/10.1145/3527188.3561937
82. Yurt, E., and Kasarci, I. (2024). A Questionnaire of Artificial Intelligence Use Motives: A contribution to investigating the connection between AI and motivation. International Journal of Technology in Education, 7(2).

83. Zhang, Y., & Gosline, R. (2023). Human favoritism, not AI aversion: People's perceptions (and bias) toward generative AI, human experts, and human-GAI collaboration in persuasive content generation. Judgment and Decision Making, 18, e41.
84. Zlateva, P., Steshina, L., Petukhov, I., and Velev, D. (2024). A conceptual framework for solving ethical issues in generative artificial intelligence. In Electronics, Communications and Networks (pp. 110-119). IOS Press.
85. Złotowski, J., Yogeeswaran, K., & Bartneck, C. (2017). Can we control it? Autonomous robots threaten human identity, uniqueness, safety, and resources. International Journal of Human-Computer Studies, 100, 48-54.

# Appendix. Summary of Attitudes to Generative AI as Work Associate (AGAWA) Scale

The AGAWA scale, consisting of four highly coherent items (α = 0.804), proved to be an accurate and reliable tool for assessing attitudes toward GenAI as a collaborator. Its strength lies in its compactness (only four items), and therefore, the short time required to complete the questionnaire and compute the score, as well as in its consideration of two dimensions: pragmatic (S1 - efficiency, S2 - competitive advantage, S4 - support at work) and moral (S7 - lack of moral resistance). On the other hand, AGAWA integrates two complementary perspectives on the measured attitude towards GenAI, i.e., examining cooperation with artificial agents from the workplace perspective (the first two statements) and from the entity's perspective within the workplace (the last two statements). This makes it a valuable tool to assess such attitudes, for instance, during employee recruitment.

It is crucial given the significant demand for this type of rapid assessment; shortly, every intellectual job is expected to involve cooperation with GenAI agents, which affects about 840 million employees (Gmyrek et al., 2025; International Labour Organization, 2024). The necessary conditions for effective collaboration include recognizing the benefits arising from such cooperation, being open to interaction with artificial agents, accepting GenAI as part of social life in the workplace (e.g., considering its opinions and decisions), and mitigating moral concerns regarding GenAI.

A positive attitude toward GenAI, measured using the AGAWA scale, showed a significant correlation with the declared frequency of using generative AI tools. This result provides empirical validation of the construct, confirming that a positive attitude measured by the AGAWA scale does indeed translate into a willingness to use GenAI more frequently. This means that attitudes toward GenAI are not merely a declaration of worldview, but are reflected in real-world behaviors, including the actual use of generative AI technology.

The AGAWA scale consists of the following four items:
1. Companies that utilize generative artificial intelligence will gain a competitive advantage over those that do not in the near future (orig. Firmy korzystające z pomocy generatywnej sztucznej inteligencji, w niedalekiej przyszłości zdobędą przewagę nad firmami nie korzystającymi z tej technologii).
2. Companies should use generative artificial intelligence as extensively as possible - of course, only within legal bounds - to achieve maximum efficiency (orig. Firmy powinny korzystać z pomocy generatywnej sztucznej inteligencji w jak największym stopniu (oczywiście tylko w ramach legalnych działań), aby osiągnąć jak największą wydajność).
3. Generative artificial intelligence will prove to be a support in my professional work. (orig. Generatywna sztuczna inteligencja okaże się wsparciem w mojej pracy zawodowej).
4. Using generative artificial intelligence as an assistance at professional work does not raise my moral objection (orig. Używanie generatywnej sztucznej inteligencji jako pomocy w pracy zawodowej nie budzi mojego moralnego oporu).

Participants respond to these items using a seven-point Likert scale ranging from 1 (complete rejection) to 7 (full acceptance). The attitude level is calculated as the arithmetic mean of the item scores, where values lower than four indicate a negative attitude toward GenAI and higher values indicate a positive one. The scale is reliable. Its validity has been confirmed by expert evaluation.